# Multi-Object Portion Tracking in 4D Fluorescence Microscopy Imagery with Deep Feature Maps

Yang Jiao[1]
jiaoy1@unlv.nevada.edu

Mo Weng[2]
mo.weng@unlv.edu

Mei Yang[1]
mei.yang@unlv.edu

[1] Department of Electrical and Computer Engineering, University of Nevada, Las Vegas
[2] School of Life Science, University of Nevada, Las Vegas

## Abstract

*3D fluorescence microscopy of living organisms has increasingly become an essential and powerful tool in biomedical research and diagnosis. An exploding amount of imaging data has been collected, whereas efficient and effective computational tools to extract information from them are still lagging behind. This is largely due to the challenges in analyzing biological data. Interesting biological structures are not only small, but are often morphologically irregular and highly dynamic. Although tracking cells in live organisms has been studied for years, existing tracking methods for cells are not effective in tracking subcellular structures, such as protein complexes, which feature in continuous morphological changes including split and merge, in addition to fast migration and complex motion. In this paper, we first define the problem of multi-object portion tracking to model the protein object tracking process. A multi-object tracking method with portion matching is proposed based on 3D segmentation results. The proposed method distills deep feature maps from deep networks, then recognizes and matches objects' portions using an extended search. Experimental results confirm that the proposed method achieves 2.96% higher on consistent tracking accuracy and 35.48% higher on event identification accuracy than the state-of-art methods.*

## 1. Introduction

Fluorescence microscopy has undergone an evolution in the past decades. Using a variety of fluorescent indicators, specific targets such as proteins, lipids, or ions can be tailored and, therefore, visualized in live [1], [2]. Rather than relying on physical sections of chemically fixed tissues, technologies such as confocal and multi-photon microscopy enable the acquisition of optical sections of thick biological objects. By excluding out-of-focus light or specifically activating fluorophores in the focus plane, a 2D image can be obtained. This allows accurate reconstruction of the 3D structures of biological samples and continuous imaging of living cells or organisms. Time-lapse movies of 3D images result in 4D fluorescence microscopy data with temporal information as the additional dimension.

The current bottleneck in biomedical research and diagnosis is to effectively extract the information from the increasingly large and complex biological image dataset in quantitative ways. In doing so, there are two types of challenges. One type is the physical limitations in image acquisition, including the number of pixels per image and signal-to-noise ratio (SNR) [3]. Besides physical size and imaging speeds of the equipment, the number of pixels per image also depends on the temporal resolution in live imaging, which in turn is decided by the time scale of the observed biological activities. Biological samples often suffer from a low SNR. There is always a limitation to how many fluorophores an object of interest can be labeled with. Besides, all fluorophores are subjected to photobleaching [4], a phenomenon where the fluorophore stops emitting fluorescence after repeated exposure to lasers. Fluorescent proteins used in live imaging are especially sensitive to photobleaching. Frequent and long-term imaging lead to a severe reduction in fluorescent signals resulting in a low SNR.

The other type of challenge comes from the intrinsically dynamic behaviors of the biological objects of interest. Living objects can often change their morphologies rapidly, which is especially true for protein machinery inside cells. Furthermore, many biological objects display constant movement and even interact with other objects of the same type. One well-studied example is tracking cells in living organisms [5]. Different from macro objects, e.g. cars or bicycles, cells may present deformations such as elongation, expansion, and shrinkage [6], as well as demonstrate complex motion patterns in a short time period [7]. However, compared to subcellular structures, which are often the machinery that perform tasks for cells, cells largely maintain constant volumes and have a nucleus that is often large and easily trackable. Subcellular structures, such as protein clusters, grow or shrink much more rapidly. They can also interact with each other through split and merge.

Conventional tracking methods used for biomedical image analysis employ either nearest neighbor linking or a motion detector, like a Kalman filter (KF) [5]. These methods can be limited in tracking fast migration and complex motions including split and merge. In recent years, deep learning methods have been applied to biomedical image analysis [8]. Due to the lack of remarkable features, object recognition in biomedical images is rather difficult compared with the detection of macro objects. It has been discovered that instead of using object position or motion path, like nearest neighbor linking and KF, tracking by object recognition is more reliable [9]. However, the physical scale and more dynamic properties of functional protein clusters make conventional cell tracking methods ineffective in producing reliable results.

In this paper, to tackle the challenges faced by cell/protein tracking, we propose a multi-object portion tracking method with portion matching for 3D fluorescence microscopy images. Based on 3D segmentation results, the proposed deep feature map tracking method distills deep feature maps from deep learning networks, then recognizes and matches objects' portions using an extended search. The proposed method conquers the challenges of rapid deformation, as well as object birth and death, and object split and merge. The method is evaluated by the 3D fluorescence images of E-Cadherin fused with Green Fluorescent Protein (GFP) from developing Drosophila (fruit fly) embryos and compared with four tracking methods.

The remainder of the paper is organized as follows: Section 2 reviews the related work in object segmentation and tracking. Section 3 defines the multi-object portion tracking problem and presents the deep feature map tracking method with portion matching. Section 4 presents the experimental results. Section five concludes the paper.

## 2. Related Work

A critical challenge in using fluorescence microscopy images is the presence of noise. Denoising filters, such as the Gaussian filter, average filter, etc., are widely employed to reduce such noises [10]–[12]. In the literature, diverse cell and nuclei segmentation methods have been studied, including active contour, watershed, and thresholding methods, as well as deep neural network methods. The active contour model, as known as snakes, delineates the outlines of an object. By means of energy minimization, active contour adapts to differences and noise, but requires a known approximation of the searching boundary [13], [14]. Via extending active contour to 3D active surface, cellular objects in sequences of microscopy images are identified [15]. The watershed method is advanced on separating touching objects [16]. By defining foreground and background locations, marker-controlled watershed segmentation acquires improved performance in nuclei segmentation [17]. Thresholding is another frequently applied segmentation method. In [18], Otsu thresholding [19] is modified to adapt with 2D microscopy image segmentation.

Recently, convolutional neural networks (CNN) [20] lead the frontier in object classification and semantic segmentation. Since 2015, several 2D semantic segmentation architectures have been proposed, including Fully Connected Networks (FCN) [21], SegNet [22], U-Net [23], and Fully Connected Dense Networks (FCDN) that adopt ResNet [24]–[26], etc. Among the aforementioned networks, U-Net demonstrates a capacity for biomedical image segmentation using a combined skip-connection that directly connects down-sampling and up-sampling layers [8]. However, stacking 2D segmentation results into 3D volume may cause a problem of misalignment in the depth. Extending to 3D data, a 3D U-Net [27] is built to learn volume information. 3D labeling is able to be generated using Generative Adversarial Networks (GAN) and used in training for 3D segmentation in terms of the labeling challenge for annotators [28].

Tracking is a complex processing, following object detection and segmentation. The popular cell or particle tracking methods can be divided into three categories: Nearest neighbor [29]–[33], Kalman filter [34]–[36] data association [34], [35], and deep learning [7], [10]. The nearest neighbor method is the simplest approach, which links every segmented object to the nearest object in the next frame [37]. It has been applied to track label-free single cells in 3D matrices [29], as well as for particle and whole cell tracking [30]. However, as shown in [29], the nearest neighbor method fails when a cell migrates fast. A tumor cell tracking platform is developed in [31] to recreate a tightly interconnected system of cancer and immune cells with 3D environmental properties. As a method proposed for multiple object tracking (MOT) [32], [33], intersection-over-union that overlaps objects in two frames, is applied to a cell tracking benchmark [38] to perform multi-cell tracking in 2D and 3D space. A Kalman filter (KF) [39] tracks an object by projecting the current state forward (in time) and estimating error covariance to obtain the a priori estimates for the next frame [7]. KF is commonly employed with data association to improve performance, e.g., maximum likelihood, and acceptance gate associated with KF [34], [35]. To improve KF estimation, local graph matching is applied after KF in the case of plant cell tracking because only cell movement is relevant [36]. Further, deep learning architectures can be used to solve tracking problems. The common method is converting the tracking task into a 2D classification, which adopts the *t*-1 state of an object as input and predicts the *t* state [7], [10].

Despite the aforementioned methods proposed for cell tracking, cell tracking or protein cluster tracking that support split and merge is rarely studied. According to [5], among 28 tracking methods reviewed, only two [40], [41] that use watershed segmentation and the nearest neighbor tracking method support split and merge. In [42], the split and merge measurements are represented by a sparse matrix and solved by a Markov chain Monte Carlo based auxiliary particle filter. However, the basic assumptions of this approach are: (1) objects are almost non-deformable; (2) the size and shape remain the same after cell events. The Markov chain Monte Carlo data association method is then proposed in [43], which adopts multiple GFP cluster split and merge tracking for 2D frames. By defining object events into five conditions, including object born, vanish, remain, merge, split, and edge, conditions of objects are measured as distance, which is then input to the method.

## 3. Methods

### 3.1. Multi-object portion tracking problem

As discussed in Section 1, one important problem in cell/protein tracking is how to track split and merge behaviors. Via breaking the whole object tracking problem into object portion tracking, any type of object relationship, including one-to-many (split) and many-to-one (merge), can be modeled as one-to-one mapping. That is, if a portion is selected as small as possible, there only exists in one or none object in a next time frame. The multi-object portion tracking problem can be modeled using a probabilistic model as follows:

Assume $\Omega$ is the collection of all tracks in time period $T$; $Y$ is the total observation; a single track is defined as,
$$\omega_j = \Omega(j)$$
where $j$ is track ID, and a single track at time $t$ is $\omega_j(t)$.

Let $O_t$ be the observation set of objects at time $t$, assume that $o_t^i$ is an object in $O_t$, and $\delta_{O_t^i}^n$ is a portion of $o_t^i$, where $i$ is the object ID of time $t$ and $n$ is the portion ID. For the history observation $O_{t-\tau}, \tau = 1,2,...T-1$, $o_{t-\tau}^{i'}$ is an object in $O_{t-\tau}$ and $\delta_{O_{t-\tau}^{i'}}^{n'}$ is a portion of $o_{t-\tau}^{i'}$, where $\tau$ is the frame gap; $i'$ is the object ID of time $t$-$\tau$ and $n'$ is the portion ID.

Assuming that $o_{t-\tau}^{i'} \in \omega_j$, as $\delta_{O_{t-\tau}^{i'}}^{n'} \in o_{t-\tau}^{i'}$, the probability of portion matching, i.e., portion $\delta_{O_t^i}^n$ is matched with object $o_{t-\tau}^{i'}$, is set as
$$P(\delta_{O_t^i}^n | o_{t-\tau}^{i'}) = \max_{n'} P(\delta_{O_t^i}^n | \delta_{O_{t-\tau}^{i'}}^{n'}) \quad (1)$$

Then for all $\tau$, the probability that $\delta_{O_t^i}^n$ is in the track of $o_{t-\tau}^{i'}$, can be formulated as
$$P\left(\delta_{O_t^i}^n \middle| \omega_j(t)\right) = \sum_\tau P(\delta_{O_t^i}^n | o_{t-\tau}^{i'}) \quad (2)$$

For all $\delta_{O_t^i}^n \in o_t^i$, one can determine that $o_t^i$ is in the track of $o_{t-\tau}^{i'}$ by

$$P(o_t^i | \omega_j(t)) = \max_n P\left(\delta_{O_t^i}^n \middle| \omega_j(t)\right) = \max_n \sum_\tau P(\delta_{O_t^i}^n | o_{t-\tau}^{i'}) \quad (3)$$

which gives
$$P(o_t^i | \omega_j(t)) = \max_n \sum_\tau \max_{n'} P(\delta_{O_t^i}^n | \delta_{O_{t-\tau}^{i'}}^{n'}) \quad (4)$$

Assuming that the object birth/death has no impact to the tracking of one existing object, the probability of one consistent tracking in $T$ can be modeled as
$$P(\omega_j | Y) = \prod_{t=1}^T P(o_t^i | \omega_j(t))$$
$$= \prod_{t=1}^T \max_n \sum_\tau \max_{n'} P(\delta_{O_t^i}^n | \delta_{O_{t-\tau}^{i'}}^{n'}) \quad (5)$$

The objective of the multi-object portion tracking problem is to maximize $P(\omega_j | Y) \ \forall \omega_j \in \Omega$. Hence, to optimize the multi-object portion tracking, one has to maximize the sum of the maximal portion matching probability among interval $\tau$, i.e.,
$$\sum_\tau \max_{n`} P(\delta_{O_t^i}^n | \delta_{O_{t-\tau}^{i'}}^{n'})$$

The optimization can be achieved by (1) reliable segmentation because false segmentation disturbs portion matching results; (2) an efficient matching approach to improving the successful matching rate.

In the proposed approach, segmentation and tracking are based on deep feature maps extracted by a deep learning architecture. The merits of deep feature maps are (1) Abundant: deep feature maps are extracted from deep learning architectures that allow multiple map layers. Taking U-Net as an example, 64 feature map layers are preserved as effective ones before pixel-wise classification, which is preferable to single contour or surface information. (2) Reliable: deep feature maps are weight matrices that are selected and optimized by deep learning architectures. Distilled by multiple encoders and decoders, weight matrices are optimized via a global training and loss function, which raises convergence of error. (3) Complex object events supportive: birth/death, split/merge are able to be identified by recognition through deep feature maps. Traditionally, nearby search and motion prediction accelerate tracking as well as improve tracking performance by error elimination. However, they are limited by high migrating speed, high object density, and complex motion models. Deep feature maps help a tracking mechanism achieve reliable object recognition. The abovementioned limitations are thus released. (4) Multidimensional: deep feature maps can associate with data of any dimension. For example, a 3D image will have a *(X, Y, Z, D)* size of feature map where *X, Y,* and *Z* are length, width, and height of the image respectively, and *D* is the depth of feature maps. (5) Calculation efficiency: although 64-layer deep feature maps are extracted by U-Net, maps with fewer layers can be employed in tracking considering hardware and time efficiency.

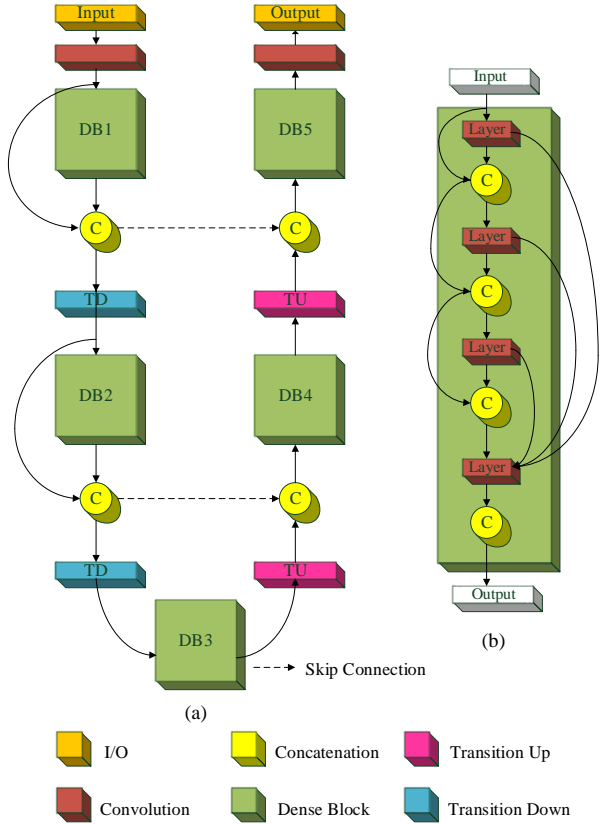

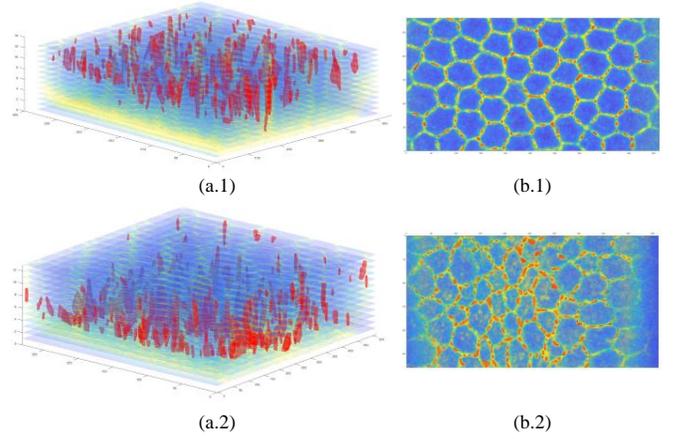

Figure 2 Segmentation results of 3D fluorescence image: (a) 3D view; (b) 2D view

Figure 1 3D asymmetric FCDN-103: (a) network architecture; (b) Dense block

Table 1 3D Asymmetric FCDN-103 Detail

|  | Layer | Transition Down | Transition Up |
|---|---|---|---|
| Sub-architecture | Batch Norm. | Batch Norm. | 3x3x3 Transposed Convolution stride=2x2x1 |
|  | ReLU | ReLU |  |
|  | 3x3x3 Convolution | 1x1x1 Convolution |  |
|  | Dropout $p$=0.2 | Dropout $p$=0.2 |  |
|  |  | 2x2x1 Max Pooling |  |

## 3.2. 3D Segmentation

Since ResNets and DenseNets present efficiency in classification, semantic segmentation architectures that adopt Residual blocks and Dense blocks have demonstrated efficiency in pixel-wise classification in recent years. Fully Convolutional DenseNets [26] (FCDN) is one of the efficient semantic segmentation architectures. Here, the 103 layered FCDN-103 is expanded into a 3D architecture to perform segmentation for 3D fluorescence microscopy images. Figure 1 shows the architecture of 3D asymmetric FCDN-103.

In [27] and [28], when building 3D deep networks, the authors employ a symmetric convolution filter and pooling filter (e.g. filter size [3, 3, 3] or [2, 2, 2]). Nevertheless, in this application, the image size ratio of the 3D fluorescence microscopy images is 280:512:13, which is asymmetric. With the resolution, image resizing may compromise the data. As such, the third dimension data is chosen to be fully preserved in the network. The layer arrangement and the other 2D parameter setting are identical as [26]. The parameter setting is shown in Table 1. Figure 2 presents the segmentation results.

## 3.3. Deep feature map tracking

The proposed deep feature map tracking (DFMT) method distills deep feature maps from the 3D segmentation results, then recognizes and matches objects' portions using an extended search.

### 3.3.1 Portions of 4D deep feature maps

After training with the Adam Optimizer and the softmax cross entropy loss function, the final parameter matrix of DB5 is $W_t^m$, which is a 64-layer 4D feature map of time $t$, and $m$ is the layer of the feature map. The 4D feature maps of all observations $O_t$ at time $t$ can be defined as

$$\Theta_t = \bigcup_{m=1}^{64} W_t^m$$

Instead of using direct image information (e.g. intensity), here, the portion matching is realized by using deep feature maps. The deep feature map portions corresponding to the object portions

$$\delta_{o_t^i}^n \in o_t^i \text{ and } \delta_{o_{t-\tau}^{i'}}^{n'} \in o_{t-\tau}^{i'}$$

are denoted as

$$\theta_{o_t^i}^n \in \Theta_t(o_t^i) \text{ and } \theta_{o_{t-\tau}^{i'}}^{n'} \in \Theta_{t-\tau}(o_{t-\tau}^{i'}),$$

respectively.

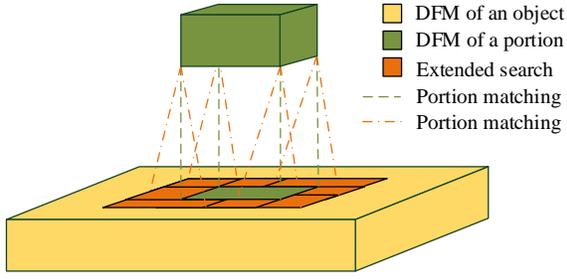

Figure 3 3D view of the 4D extended search and portion matching using deep feature map (DFM)

### 3.3.2 Extended search and portion matching

Rather than only matching between a pair of portions from two time frames, an extended search employs a set of portions to cover the surroundings in volume. It is applied to $\Theta_t$ to accommodate motions and fluctuations.

Taking a pixel $p(x_p, y_p, z_p) \in o_t^i$ as the center point in $x$, $y$, and $z$ dimension of object $o_t^i$, the deep feature map portion is defined as

$$\theta_{o_t^i}^n = \{\Theta_t(x_p \pm r_x, x_p \pm r_y, x_p \pm r_z, m), \forall m\}$$

where $2r_x, 2r_y$ and $2r_z$ are the lengths in dimensions $x$, $y$, and $z$ of each portion, respectively. For the extended search in frame $t - \tau$, a pixel group is obtained by

$$p_\tau \in \{(x_{p\tau}, y_{p\tau}, z_{p\tau}) \mid x_{p\tau} \in (x_p \pm r_x^{ext}), y_{p\tau} \in (y_p \pm r_y^{ext}), z_{p\tau} \in (z_p \pm r_z^{ext})\}$$

where $r_x^{ext}, r_y^{ext}, and\ r_z^{ext}$ are the extended range in $x$, $y$, and $z$ dimension, respectively. Then an object's deep feature map portion in frame $t - \tau$ is defined as $\forall p_\tau(x_{p\tau}, y_{p\tau}, z_{p\tau})$,

$$\theta_{o_{t-\tau}^{i'}}^{n'} = \{\Theta_{t-\tau}(x_{p\tau} \pm r_x, x_{p\tau} \pm r_y, x_{p\tau} \pm r_z, m), \forall m\}$$

The portion matching between $\theta_{o_t^i}^n$ and $\theta_{o_{t-\tau}^{i'}}^{n'}$ is defined by Pearson's correlation coefficient as

$$\rho\left(\theta_{o_t^i}^n\right) = \frac{\sum_{x,y,z,n}(\theta_{o_t^i}^n(x,y,z,m) - E(\theta_{o_t^i}^n))(\theta_{o_{t-\tau}^{i'}}^{n'}(x,y,z,m) - E(\theta_{o_{t-\tau}^{i'}}^{n'}))}{\sqrt{\sum_{x,y,z,n}(\theta_{o_t^i}^n(x,y,z,m) - E(\theta_{o_t^i}^n))^2(\theta_{o_{t-\tau}^{i'}}^{n'}(x,y,z,m) - E(\theta_{o_{t-\tau}^{i'}}^{n'}))^2}} \quad (6)$$

where $E(\theta_{o_t^i}^n)$ is the expectation of the portion $\theta_{o_t^i}^n$. If a portion $\theta_{o_t^i}^n$ is associated with $l$ objects in the extended search space among $\tau$ time frames $\{O_t^{i_1}, O_t^{i_2}, \ldots O_{t-\tau}^{i_l}\}$, it has an array of coefficient $\boldsymbol{\rho}\left(\boldsymbol{\theta_{o_t^i}^n}\right)$ describing its correlation with all $l$ objects. The best matched object portion is selected by

$$M\left(\theta_{o_t^i}^n\right) = \left\{k \mid \rho\left(\theta_{o_t^k}^n\right) = \max_{i=i_1 \to i_l} \boldsymbol{\rho}\left(\boldsymbol{\theta_{o_t^i}^n}\right) \geq \gamma\right\} \quad (7)$$

where $\gamma$ is a lower bound of acceptance for portion matching. $M\left(\theta_{o_t^i}^n\right)$ is a set of matched object ID. Figure 3 shows a 3D view of the 4D extended search and portion matching.

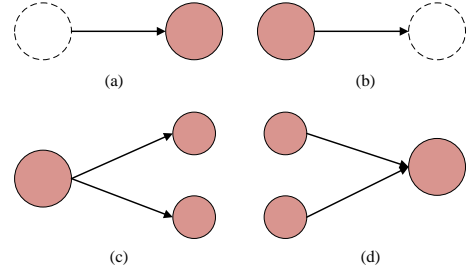

Figure 4 Object events (a) birth (b) death (c) split (d) merge

### 3.3.3 Object events identification

Besides one-to-one object mapping, the object events of cells and protein clusters include four types: birth, death, split, and merge, as shown in Figure 4.

**Birth:** An object $O_t^i$ is newborn if none of its portions are matched with any object portion in $\tau$ former time frames, i.e.,

$$\cup_n M\left(\theta_{o_t^i}^n\right) = \emptyset, \forall n. \quad (8)$$

New IDs are assigned to newborn objects.

**Death:** Observations for dead objects at time point $t$ are not available. These objects' IDs will be held and never assigned to other objects in case of reappearance.

**Split:** Two or more objects are considered split from one parent if the intersection of unions of the matched object ID set is not empty. New object IDs are assigned to child objects but parents' IDs are recorded for tracking.

$$\cup_n M\left(\theta_{o_t^i}^n\right) \cap \cup_{n'} M\left(\theta_{o_t^{i'}}^{n'}\right) \cap \ldots \neq \emptyset, \forall n, n', \ldots \quad (9)$$

**Merge:** An object is considered merged if the matched object union contains multiple elements as

$$\cup_n M\left(\theta_{o_t^i}^n\right) = \{i_1, i_2, \ldots\}. \quad (10)$$

The object ID of the merged object is inherited from the major (larger size) parent object, though all parents' IDs are recorded.

Via DFMT, the matching relationships between portions and corresponding objects are identified. The collection of all matched objects at time $t$ is defined as

$$\cup_i \cup_n M\left(\theta_{o_t^i}^n\right) \quad (11)$$

Figure 5 plots the sample correlation map of protein objects in the time interval [$t$, $t$-$\tau$]. Through portion matching with extended search in $\tau$ time frames, DFMT successfully identifies the split and merge relations among multiple objects. Object IDs are inherited from the parents following the matching relationships.

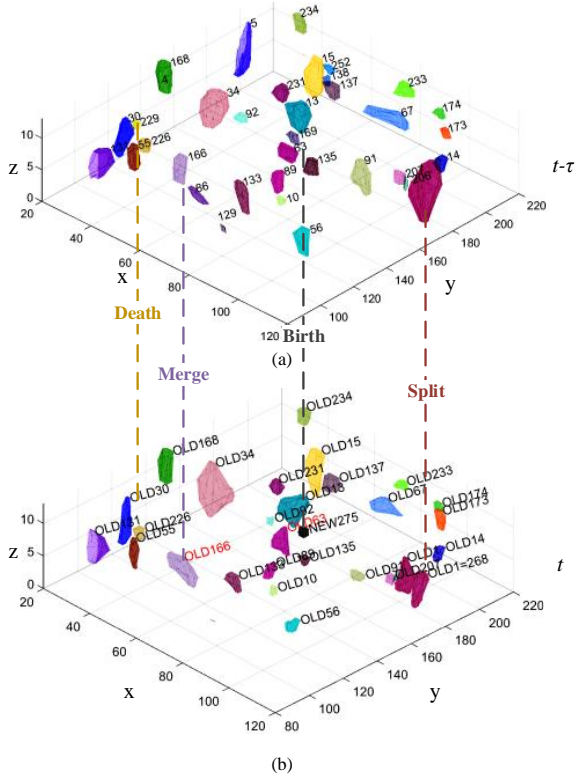

Figure 5 Correlation maps. Each object is in a unique color.

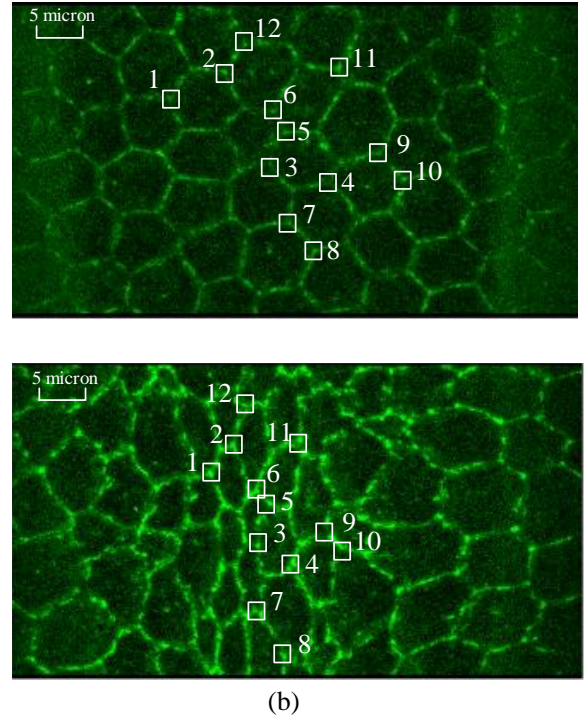

(b)
Figure 6 Ground truth objects (a) t=1 (b) t=69

## 4. Experimental Results

### 4.1. Dataset

The 4D dataset used for evaluating the proposed method is a time-lapse movie taken from developing Drosophila (fruit fly) embryos. The surface of each embryo is covered by a single layer of columnar-shaped epithelial cells, with the top of the cells on the surface of the embryo. The dimension of the columnar cell is proximately 6.5 micron wide and 30 microns tall. E-Cadherin fused with Green Fluorescent Protein (GFP) is expressed in the entire embryo to visualize adherent junctions, the major cell-to-cell junctions that physically connect neighbor cells. E-Cadherin is the membrane component of adherent junctions and therefore localize exclusively on the cell membrane. At this point of development, E-Cadherin molecules are organized into small clusters of various sizes on the lateral membrane of the columnar cells.

Images were taken on a Leica SP5 laser scanning confocal microscope. The 488 μm laser was used to excite the GFP. At each time point, a 60-micron by 30-micron area was imaged and a z-stack of optical sections were taken from the top of the cell to six microns below, creating a 3-D data set. The developing embryo was imaged this way every five seconds, generating a 4-D data set.

### 4.2. Evaluation

Based on the same segmentation result generated by the asymmetric 3D FDCN in Section 3.1, seven tracking methods, including the proposed DFMT method, are tested and compared with the ground truth. For the comparing methods, Method1 (M1) [32] employs intersection-over-union (IOU), which scores the cell relations by Eq.12

$$IOU(a,b) = \frac{Area(a) \cap Area(b)}{Area(a) \cup Area(b)} \quad (12)$$

where $a$ and $b$ are two objects. Based on the performance, the threshold $\sigma_{IOU}$ is set to $0$ to maximize the tracking result. Method2 (M2) [43] uses object linking and expands linking (10 pixels) when objects move. Besides, overtime tracking is applied in Method2. For object events, a split is determined if two objects $a$ and $b$ in frame $t$ are linked to one object $c$ in frame $t$-1, and merge is determined if object $a$ in frame $t$ has two sources $b$ and $c$ in frame $t$-1 when $size(a) \geq 1.5 \times average\ size(b, c)$. Since Method1 does not support object event identification, Method1 is evaluated one more time with the object event identification approach of Method2, shown as M1+M2. Method3 (M3) [44] is applied through Fiji [45] and ImageJ [46]. It supports tracking with (M3 (S&M)) or without (M3) split and merge detection. Both versions are tested.

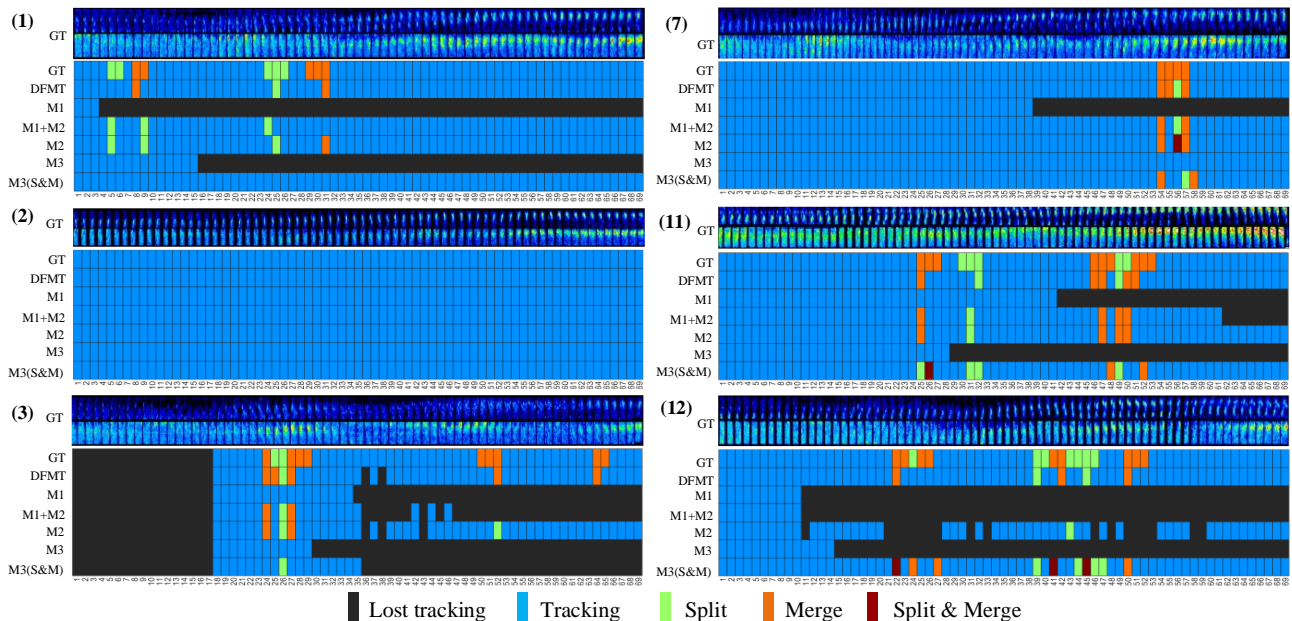

| | Lost tracking | | Tracking | | Split | | Merge | | Split & Merge |

Figure 7 Object tracking and event identification results

Table 2 Object tracking and event identification results

| Methods | Tracking Accuracy | Split Accuracy | Merge Accuracy | S&M Accuracy | S&M Sensitivity | S&M Precision |
|---|---|---|---|---|---|---|
| **DFMT** (proposed) | **98.52%** | **81.82%** | **85.00%** | **83.87%** | **83.87%** | **96.30%** |
| M1 [32] | 43.28% | - | - | - | - | - |
| M1+M2 [32], [41] | 77.31% | 63.64% | 40.00% | 48.39% | 50.00% | 83.33% |
| M2 [41] | 95.56% | 54.55% | 40.00% | 45.16% | 53.85% | 82.35% |
| M3 [44] | 65.84% | - | - | - | - | - |
| M3 (S&M) [44] | 93.09% | 63.64% | 40.00% | 48.39% | 51.72% | 88.24% |

As Figure 6 shows, the evaluation is based on the tracking of 12 objects through 69 time points. Due to the difficulty of determining a single time point of object events by the naked eye, a time range is set for each object event. Figure 7 presents the ground truth of split and merge events in the time range. For evaluation purposes, if a method detects a respond object event within the time range, the detection is positive. Any missed detection is determined as false negative (FN), and any false detection is counted as false positive (FP). The evaluation metrics include accuracy, sensitivity, and precision, as defined by the following equations.

$$\text{Precision} = TP/(TP + FP) \quad (13)$$

$$\text{Sensitivity} = TP/(TP + FN) \quad (14)$$

$$\text{Accuracy} = (TP + TN)/(TP + FP + TN + FN) \quad (15)$$

Figure 7 demonstrates the tracking of six example objects. In Table 2, tracking accuracy describes the capacity of consistent tracking in time period *T*. The split/merge accuracy represents the accuracy of split/merge event identification. Split and merge event accuracy, sensitivity, and precision show the comprehensive performance for both split and merge events. From Figure 7 and Table 2, one can see that without split and merge detection, M1 and M3 lose tracking with higher frequency. The proposed DFMT method achieves long-term tracking at a highly successful rate, among all methods. Besides, it significantly improves the performance of object event identification. Both the split and merge of protein clusters are accurately detected with low FN and FP. The proposed method achieves 2.96% higher on consistent tracking accuracy than Method2 and 35.48% higher on event identification accuracy than Method3, respectively. The efficiency of DFMT is delivered by the valid recognition of objects based on the accuracy and reliability of the deep feature maps. It is confirmed that the deep feature map tracking method is powerful in cases including fast migration, dense distribution, and complex motion models. Figure 8 shows the 3D tracking result by the proposed method.

## 5. Conclusion

In this paper, we proposed the deep feature map tracking method for tracking objects in 4D fluorescence imagery. The DFMT method targets to complex motion objects, such as protein clusters, which may demonstrate dense distribution, fast migration, rapid volume alternation, split, and merge. The proposed method works in two main steps. First, based on the 3D segmentation result, the 4D deep feature map is extracted from the last dense block for each 3D fluorescence frame. Next, portion matching between objects in two frames is performed by finding the highest correlation within the extended search space. Consequently, partial or whole object tracking, covering object events like birth, death, split, and merge is realized by the proposed method. The experiment results demonstrate that the DFMT method achieves a high success rate on long-term tracking and object event identification.

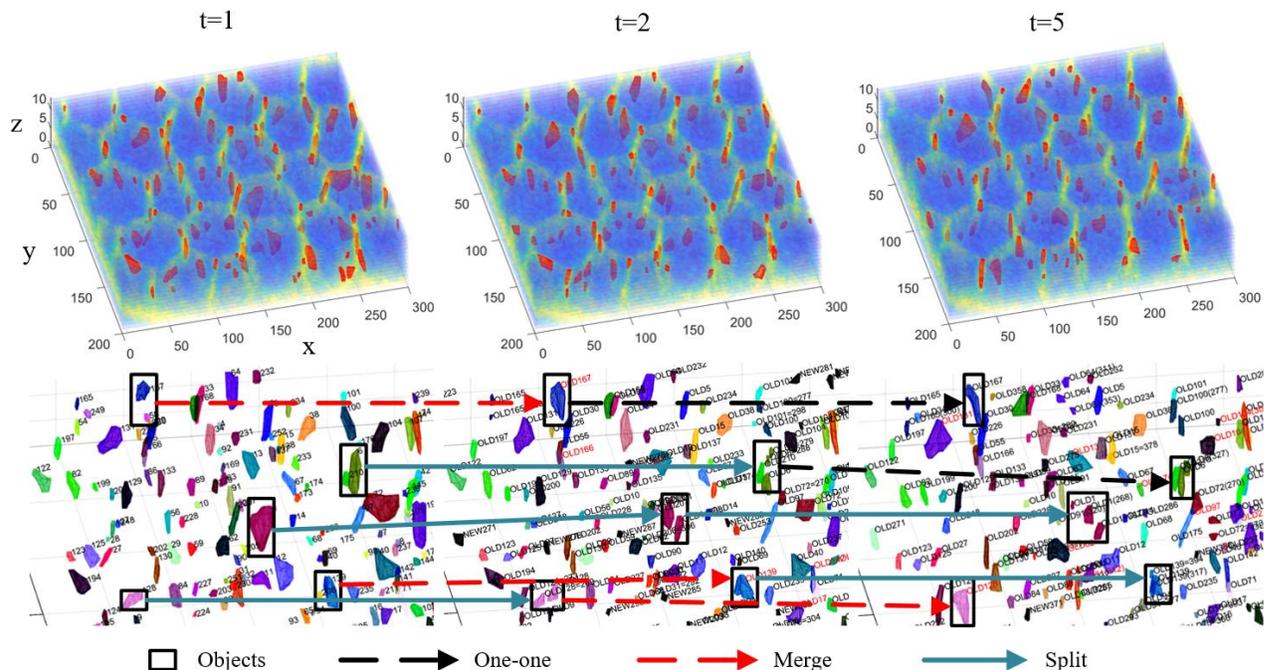

Figure 8 3D tracking by the proposed approach. Object IDs and colors are inherited if tracking is successful.

## Acknowledgements

We acknowledge the support for this research: UNLV TTGRA and NIH Pathway to Independence Award (K99/R00).